\documentstyle[12pt]{article}
\newcommand{\bce}{\begin{center}}
\newcommand{\ece}{\end{center}}
\newcommand{\beq}{\begin{equation}}
\newcommand{\eeq}{\end{equation}}
\newcommand{\bea}{\vspace{0.25cm}\begin{eqnarray}}
\newcommand{\eea}{\end{eqnarray}}

\newcommand{\ba}{\begin{array}}
\newcommand{\ea}{\end{array}}

\newcommand{\doublespace}{
    \renewcommand{\baselinestretch}{1.6}\large\normalsize}

\def\lsim{\mathrel{\rlap{\lower4pt\hbox{\hskip1pt$\sim$}}
    \raise1pt\hbox{$<$}}}	  %less than or approx. symbol
\def\gsim{\mathrel{\rlap{\lower4pt\hbox{\hskip1pt$\sim$}}
    \raise1pt\hbox{$>$}}}	  %greater than or approx. symbol

\def\Pom{{\bf I\!P}}

\def\lsim{\mathrel{\rlap{\lower4pt\hbox{\hskip1pt$\sim$}}
    \raise1pt\hbox{$<$}}}         %less than or approx. symbol
\def\gsim{\mathrel{\rlap{\lower4pt\hbox{\hskip1pt$\sim$}}
    \raise1pt\hbox{$>$}}}         %greater than or approx. symbol

\def\Pom{{\bf I\!P}}

  \advance\oddsidemargin  by -1in
  \advance\evensidemargin by -1in
  \advance\topmargin      by -0.5in
\def\lsim{\mathrel{\rlap{\lower4pt\hbox{\hskip1pt$\sim$}}
    \raise1pt\hbox{$<$}}}         %less than or approx. symbol
\def\gsim{\mathrel{\rlap{\lower4pt\hbox{\hskip1pt$\sim$}}
    \raise1pt\hbox{$>$}}}         %greater than or approx. symbol

\def\Pom{{\bf I\!P}}
\def\beq{\begin{equation}}
\def\endeq{\end{equation}}
\def\arr{\begin{eqnarray}}
\def\endarr{\end{eqnarray}}
\makeindex

\begin{document}

\phantom{.}\hspace{8.3cm}{\large \bf KFA-IKP(TH)-1995-07}  \\
\phantom{.}\hspace{9.5cm}{\large \bf 15 May 1995}
%\vspace{1.5cm}\\
\begin{center}
{\LARGE Exploratory study of
shrinkage of the \bigskip\\
diffraction cone
 for the generalized
BFKL pomeron}
\vspace{1.0cm}\\
{\large \bf
N.N.~Nikolaev$^{a,b}$, B.G.~Zakharov$^{b,c}$ and V.R.Zoller$^{d}$
\bigskip\\}
{\it
$^{a}$IKP(Theorie), KFA J{\"u}lich, 5170 J{\"u}lich, Germany
\medskip\\
$^{b}$L. D. Landau Institute for Theoretical Physics, GSP-1,
117940, \\
ul. Kosygina 2, Moscow 117334, Russia.\medskip\\
$^{c}$ Interdisciplinary Laboratory, SISSA, via Beirut, I-34014,
Theoretical Physics,\\
Trieste, Italy. \medskip\\
$^{d}$ Institute for Theoretical and Experimental Physics,\\
Bolshaya Cheremushkinskaya 25, 117259 Moscow, Russia.
}
\vspace{1.5cm}\\
{\large \bf Abstract
\bigskip\\ }
\begin{minipage}[h]{14cm}

In color dipole gBFKL dynamics, we describe the emerging
gBFKL phenomenology of a subasymptotic energy dependence of the
diffraction slope and discuss possibilities of testing the gBFKL
predictions in exclusive photo- and electroproduction of vector
mesons $V$ at HERA. A substantial shrinkage of the
diffraction cone
$\gamma^{*}p\rightarrow Vp$ processes from the CERN/FNAL to
HERA range of energy $W$ is predicted. This subasymptotic
shrinkage
is faster than expected from  the small slope of pomeron's Regge
trajectory $\alpha_{\Pom}'$. We point out that the
diffraction slope is a scaling function of $(m_{V}^{2}+Q^{2})$,
what relates production of different vector mesons.
\end{minipage}
\end{center}
\pagebreak

\doublespace
Is the QCD pomeron a fixed or moving singularity? Can the two
options be distinguished experimentally in hard diffraction
processes at HERA? These pressing issues are adressed in
this paper to.

The
early discussion on the BFKL (Balitsky-Fadin-Kuraev-Lipatov [1])
pomeron centered on the scaling $\alpha_{S}=const$ approximation with
infinite propagation radius $R_{c}$
of (massless) gluons. In this approximation, for
a lack of the dimensional scale, the BFKL pomeron is a fixed cut
in the $j$-plane: $-\infty < j \leq \alpha_{\Pom}(t) =
\alpha_{\Pom}(0) = 1+\Delta_{\Pom}$, which also implies the
energy-independent diffraction slope at $t=0$. The scaling
approximation is not self-contained, though, because the diffusion
BFKL Green's function makes high-energy behavior
even at short distances increasingly sensitive to the nonperturbative
large-distance contribution [1-3]. The introduction of
running QCD coupling $\alpha_{S}(r)$ into the generalized BFKL
(gBFKL) equation for color dipole cross section [3]
further amplifies an intrusion of large-distance effects.
Furthermore, in our Ref.~[4] we have shown that
breaking of scale invariance by a running
$\alpha_{S}(r)$ coplemented by
the finite gluon propagation radius $R_{c}$ [3], profoundly changes
the very nature of the gBFKL pomeron from a {\sl fixed} cut
of the scaling $\alpha_{S}=const$ approximation to a {\sl moving}
cut with the finite Regge slope $\alpha_{\Pom}'$ of the pomeron
trajectory.
This conclusion [4] derives
from the fact that gBFKL diffraction
slope $B(\xi,r)$ for scattering of a color dipole of size $r$
has the asymptotic Regge growth
\beq
B(\xi,r)=B(0,r)+2\alpha'_{\Pom}\xi \, ,
\label{eq:1}
\endeq
where $\xi=\log(W^{2}/s_{o})$, $W$ is the c.m.s. energy and
$s_{0}=m_{p}^{2}$ in hadronic scattering,
$s_{0}=Q^{2}$ and $\xi=\log({1\over x})$ in deep inelastic scattering
and in leptoproduction of vector mesons $s_{0}=m_{V}^{2}+Q^{2}$.
In the gBFKL dynamics [3-5] a dimensionful
$\alpha'_{\Pom}$ is a noneperturbative quantity
related to the nonperturbative infrared
parameter of the model - the gluon propagation radius $R_{c}$.
For the preferred $R_{c}= 0.27$\,fm, quite a small
$\alpha_{\Pom}'= 0.072$\,GeV$^{-2}$ is found
(the Regge phenomenology of soft hadronic scattering suggests
\footnote{Here one must bear in mind that because of the so-called
absorption corrections the observed $\alpha_{soft}'$ is
typically larger that the Regge slope $\alpha_{eff}'$ in the
bare pomeron amplitude [7].}
$\alpha_{soft}' \sim  0.2$\,GeV$^{-2}$ [6,7]).
A strong argument [3-5,8-10]
for using running $\alpha_{S}(r)$ is that
only in this case the short-distance (Leading-Log$Q^{2}$) limit of
the gBFKL equation matches the GLDAP equation [11]. A dramatic impact
of scale invariance breaking on
the intercept $\Delta_{\Pom}$ found in [3] (for earlier work
see [9]) has recently been confirmed in [10,12]
using a very different technique.

How the finding that the gBFKL pomeron
is a moving cut can be tested experimentally?
Based on a solution
of the gBFKL equation [4] for $B(\xi,r)$, in this paper we
discuss
the emerging gBFKL phenomenology of the diffraction slope.
First, with plausible and partly tested boundary conditions,
we show that
at attainable subasymptotic energies (comprising
the HERA energies), the local Regge slope $\alpha_{eff}'(\xi,r)=
{1\over 2}\partial B(\xi,r)/\partial \xi$ is much larger
than the avove cited $\alpha'_{\Pom}$ and is
surprisingly close to $\alpha_{soft}'\sim 0.2$\,GeV$^{-2}$. Second,
we find $B(\xi,r)$ of a magnitude which is intriguingly
close to the experimental determinations. Third, we discuss
scaling properties of
$B(\xi,r)$ which relate diffraction slope for different exclusive
reactions $\gamma^{*}p\rightarrow Vp$ ($V=\rho^{0},
\omega^{0},\phi^{0},J/\Psi,\Upsilon,..$) and can be tested
at HERA. Here we focus on the
fact [13,14] that exclusive vector meson production
probes the dipole cross section
$\sigma(\xi,r)$, and $B(\xi,r)$ and $\alpha_{eff}'(\xi,r)$ thereof,
at $r\sim r_{S}$, where the scanning radius
\beq
r_{S} \approx {6\over \sqrt{Q^{2}+m_{V}^{2}}} \, .
\label{eq:2}
\endeq

We recall first
the gBFKL equation [3,5] for color dipole cross
section $\sigma(\xi,r)$:
\arr
{\partial \sigma(\xi,r) \over \partial \xi} ={\cal K}\otimes
\sigma(r,\xi)=~~~~~~~~~~~~~~~~~~~~~~~~~~~~~~~~~~~~~~~~
\nonumber\\ {3 \over 8\pi^{3}} \int d^{2}\vec{\rho}_{1}\,\,
\mu_{G}^{2}
\left|g_{S}(R_{1})
K_{1}(\mu_{G}\rho_{1}){\vec{\rho}_{1}\over \rho_{1}}
-g_{S}(R_{2})
K_{1}(\mu_{G}\rho_{2}){\vec{\rho}_{2} \over \rho_{2}}\right|^{2}
[\sigma(\xi,\rho_{1})+
\sigma(\xi,\rho_{2})-\sigma(\xi,r)]   \, \, .
\label{eq:3}
\endarr
Here the kernel ${\cal K}$ is related to the flux of
Weizs\"acker-Williams (WW) soft gluons generated by the $\bar{q}$-$q$
color dipole source,
$\vec{r}$ is the $\bar{q}$-$q$ separation
and $\vec{\rho}_{1,2}$ are the $q$-$g$ and $\bar{q}$-$g$ separations
in the two-dimensional impact parameter plane,
$K_{\nu}(x)$ is the modified Bessel function,
$\vec{{\cal E}}(\vec{\rho})= \mu_{G}g_{S}(\rho)K_{1}(\mu_{G}\rho)
{\vec{\rho} \over \rho}= -g_{S}(\rho) \vec{\nabla}_{\rho}
K_{0}(\mu_{G}\rho)$ describes a Yukawa screened
transverse chromoelectric field of the relativistic quark and
\beq
\mu_{G}^{2}
\left|g_{S}(R_{1})
K_{1}(\mu_{G}\rho_{1}){\vec{\rho}_{1}\over \rho_{1}}
-g_{S}(R_{2})
K_{1}(\mu_{G}\rho_{2}){\vec{\rho}_{2} \over \rho_{2}}\right|^{2}
=
|\vec{{\cal E}}(\vec{\rho}_{1}) - \vec{{\cal E}}(\vec{\rho}_{2})|^{2}
\label{eq:4}
\endeq
gives a flux (a modulus of the Poynting vector) of WW soft
gluons in the $q\bar{q}g$ state and
${9\over 8}[\sigma(\xi,\rho_{1})+\sigma(\xi,\rho_{2})-\sigma(\xi,r)]$
is a change of cross section for the presence of
the WW gluon [5].
The running QCD charge
$g_{S}(r)=\sqrt{4\pi \alpha_{S}(r)}$
must be taken at the shortest relevant
distance $R_{i}={\rm min}\{r,\rho_{i}\}$ and
tn the numerical analysis
[3]  an infrared freezing $\alpha_{S}(q^{2}) \leq 0.82$
has been imposed
on the three-flavor, one-loop $\alpha_{S}(q^{2})=
4\pi/[9\log(q^{2}/\Lambda^{2})]$ with $\Lambda=0.3$\,GeV. The
preferred choice $R_{c}=0.27$\,fm  gives
$\Delta_{\Pom}=0.4$ [4] and leads to a very good
description [15] of the HERA data on the proton structure function
at small $x$.

The nonperturbative infrared parameter
$R_{c}=1/\mu_{G}$ has a very clear meaning of a correlation
(propagation, Ukawa screening)
radius for perturbative gluons. The above infrared
regularization is not unique,
but the crucial transversality property of WW gluons
holds independent of the specific model for screening and only the
numerical results can somewhat change. At $r,\rho_{1,2} \ll R_{c}$ and
in the $\alpha_{S}=const$ approximation, the scaling BFKL equation is
obtained ([3,5], see also Mueller and Patel [16]), in the
Leading-Log$Q^{2}$ regime of $r\ll \rho_{1,2} \ll R_{c}$, the gBFKL
Eq.~(\ref{eq:3}) matches [3,5] the GLDAP equation
(see also [9,10]).

Generalization of (\ref{eq:3}) to the equation for
diffraction slope
$B(\xi,r)$ proceeds as follows [4]:
If the impact-parameter representation,
$\sigma(\xi,r)=2\int d^2\vec b\, \Gamma(\xi,r,\vec{b})
$ and the diffraction slope $B(\xi,r)$ at $t=0$ equals
$B(\xi,r)= {1\over 2}\langle \vec{b}\,^{2}\rangle =
\lambda(\xi,r)/\sigma(\xi,r)$,
where
\beq
\lambda(\xi,r)=\int d^2\vec{b}~  \vec{b}\,^2~\Gamma(\xi,r,\vec{b}) \, ,
\label{eq:5}
\endeq
$\Gamma(\xi,r,\vec{b})$ is the profile function and
$\vec{b}$ is the impact parameter defined with respect
to the center of the $q$-$\bar{q}$ dipole. In the $q\bar{q}g$
state, the $qg$ and $\bar{q}g$ dipoles
have the impact parameter $\vec{b}+{1\over 2}\vec{\rho}_{2,1}$
and, undoing the impact parameter integrations in (\ref{eq:3}),
one finds:
\arr
{\partial \Gamma(\xi,r,\vec{b})\over \partial \xi} =
{\cal K}\otimes \Gamma(\xi,r,\vec{b})
%~~~~~~~~~~~~~~~~~~~~~~~~~
%\nonumber\\
={3 \over 8\pi^{3}} \int d^{2}\vec{\rho}_{1}\,\,
\mu_{G}^{2}
\left|g_{S}(R_{1})
K_{1}(\mu_{G}\rho_{1}){\vec{\rho}_{1}\over \rho_{1}}
-g_{S}(R_{2})
K_{1}(\mu_{G}\rho_{2}){\vec{\rho}_{2} \over \rho_{2}}\right|^{2}
\nonumber\\
\times
[\Gamma(\xi,\rho_{1},\vec{b}+{1\over 2}\vec{\rho}_{2}) +
\Gamma(\xi,\rho_{2},\vec{b}+{1\over 2}\vec{\rho}_{1}) -
\Gamma(\xi,r,\vec{b})]  \, .~~~~~~~~~~~~~
\label{eq:6}
\endarr
It is convenient to
separate from $B(\xi,r)$ a purely geometrical component
${1\over 8}r^{2}$ due to the color dipole's elastic form factor
and consider
$
\eta(\xi,r)=\lambda(\xi,r)-{1\over 8}r^{2}\sigma(\xi,r) \, .
$
Then, the
calculation of the moment (\ref{eq:5}) in (\ref{eq:6}) leads to
an inhomogeneous equation
\arr
{\partial \eta(\xi,r) \over \partial \xi} =
{3 \over 8\pi^{3}} \int d^{2}\vec{\rho}_{1}\,\,
\mu_{G}^{2}
\left|g_{S}(R_{1})
K_{1}(\mu_{G}\rho_{1}){\vec{\rho}_{1}\over \rho_{1}}
-g_{S}(R_{2})
K_{1}(\mu_{G}\rho_{2}){\vec{\rho}_{2} \over \rho_{2}}\right|^{2}
\nonumber\\
\times\left\{\eta(\xi,\rho_{1})+
\eta(\xi,\rho_{2})-\eta(\xi,r)+
{1\over 8}(\rho_{1}^{2}+\rho_{2}^{2}-r^{2})[
\sigma(\rho_2,\xi) + \sigma(\rho_1,\xi)]\right\}\nonumber\\
{}~~~~~~~~~~~~~~~~~~~~~~~
={\cal K}\otimes \eta(\xi,r) +\beta(\xi,r)
 \, ,
\label{eq:7}
\endarr
\arr
\beta(\xi,r) = {\cal L}\otimes \sigma(\xi,r)
%~~~~~~~~~~~~~~~~~~~~~~~~~
%\nonumber\\
={3 \over 64\pi^{3}} \int d^{2}\vec{\rho}_{1}\,\,
\mu_{G}^{2}
\left|g_{S}(R_{1})
K_{1}(\mu_{G}\rho_{1}){\vec{\rho}_{1}\over \rho_{1}}
-g_{S}(R_{2})
K_{1}(\mu_{G}\rho_{2}){\vec{\rho}_{2} \over \rho_{2}}
\right|^{2} \nonumber\\
\times(\rho_{1}^{2}+\rho_{2}^{2}-r^{2})[
\sigma(\rho_2,\xi) + \sigma(\rho_1,\xi)]
\, \, . ~~~~~~~~~~~~~~~~~~~~~~
\label{eq:8}
\endarr
A homogeneous Eq.~(\ref{eq:7}) coincides with Eq.~(\ref{eq:3}),
which has several consequences. First, if $\sigma_{1}(\xi,r)$ and
$\eta_{1}(\xi,r)$ are solutions of Eqs.~(\ref{eq:3}),(\ref{eq:7}),
then $\eta_{2}(\xi,r)=\eta_{1}(\xi,r)+\Delta b\cdot \sigma_{1}(\xi,r)$
with $\Delta b=const$ also is a solution of Eq.~(\ref{eq:7}) with
$B_{2}(\xi,r)=B_{1}(\xi,r)+\Delta b$. Second, the Regge growth of
$B(\xi,r)$ can be driven only
by the inhomegeneous term $\beta(\xi,r)$.

The detailed arguments for $\alpha'_{\Pom}\neq 0$, based on
properties of eigenvalues of the gBFKL equation (\ref{eq:3})
and
of the kernel ${\cal L}$ have been presented in [4] and need not
be repeated here. We only cite the order of magnitude estimate
\beq
\alpha_{\Pom}' \sim
{3 \over 16\pi^{2}} \int d^{2}\vec{r}\,\,\alpha_{S}(r)
\mu_{G}^{2}r^{2}
K_{1}^{2}(\mu_{G}r)  \sim
{3 \over 64\pi}R_{c}^{2}\alpha_{S}(R_{c})
\, ,
\label{eq:9}
\endeq
which clearly shows the connection between the dimensionful
$\alpha_{\Pom}'$ and the nonperturbative infrared parameter
$R_{c}$. We determine $\alpha_{\Pom}'$ from the
numerical solution of Eqs.~(\ref{eq:3},\ref{eq:7},\ref{eq:8})
as the $\xi \rightarrow \infty$ limit of $\alpha'_{eff}(\xi,r)$.
The resulting $R_{c}$ dependence of $\alpha_{\Pom}'$ is shown
in Fig.~1 and is very steep in the opposite to a
weak $R_{c}$-dependence of the intercept $\Delta_{\Pom}$ found
in [3].

The gBFKL dipole cross section $\sigma(\xi,r)$
sums the Leading-Log$(W^{2})$ multigluon production cross
sections. Consequently, as a
realistic boundary condition for the
gBFKL dynamics one can take the
two-gluon exchange Born amplitude
\arr
{\rm Im}A(\xi_{0},r,\vec{q})=
{16\pi \over 9}\int d^{2}\vec{k}\,{ \alpha_{S}(\vec{k}^{2})
\alpha_{S}(\kappa^{2})
\over
[(\vec{k}+{1\over 2}\vec{q})^{2}+\mu_{G}^{2}]
[(\vec{k}-{1\over 2}\vec{q})^{2}+\mu_{G}^{2}]}
\nonumber\\
\times[\cos({1\over 2}\vec{q}\vec{r})-\cos(\vec{k}\vec{r})]\cdot
[G_{1}(\vec{q})-G_{2}(\vec{k}+{1\over 2}\vec{q},
-\vec{k}+{1\over 2}\vec{q})] \, .
\label{eq:10}
\endarr
Here we use the normalization
${\rm Im}A(\xi_{0},r,\vec{q}=0)=\sigma(\xi_{0},r)$,
$\kappa^{2}={\rm min}\{\vec{k}^{2},C^{2}r^{-2}\})$,
$C\sim 1.5$ [17], $\vec{k}\pm {1\over 2}\vec{q}$ are the
momenta of exchanged gluons, $G_{1}(\vec{q})$ and
$G_{2}(\vec{k}_{1},\vec{k}_{2})$ are the single- and two-quark
form factors of the proton probed by gluons. The former is
customarily approximated by, and the latter in simple quark models
can be related to, the charge form factor of the proton
(for instance, see [18,19]). A very good quantitative
gBFKL description
of the HERA data on the small-$x$ proton structure functions is
obtained [15] if the Born cross section
Eq.~(\ref{eq:10}) with $R_{c}=0.27$\,fm is taken as a boundary
condition for the gBFKL equation (\ref{eq:3}) at the Bjorken variable
$x_{0}=0.03$, i.e., at $\xi_{0}=\log{1\over x_{0}}=3.5$.
Hereafter we only consider $R_{c}=0.27$\,fm.

The Born approximation for $B(\xi_{0},r)$, shown in Fig.~2, has
nice properties which admit a simple interpretation. First,
for $r \lsim R_{c}$, the Born
amplitude (\ref{eq:10}) is dominated by
perturbative $k^{2} \sim r^{-2} \gsim \mu_{G}^{2}$ for which
$G_{2}(\vec{k},-\vec{k})$ vanishes. Consequently,
$A(\xi,r,\vec{q}) \propto G_{1}(\vec{q})$ and
$B(\xi_{0},r\ll R_{c})= {1\over 3}\langle R_{1}^{2} \rangle_{p}$,
which is a generic perturbative result and holds beyond
the above simplified derivation. Second,
the $r$ dependence of the Born approximation
$B(\xi_{0},r)$
can be cast in a very symmetric and intuitively appealing form
\beq
B(\xi_{0},r)= {1\over 3}\langle R_{1}^{2} \rangle + {1\over 8}r^{2}
+\delta B =
{1\over 2}\langle r_{cms}^{2}\rangle_{beam}
+{1\over 2}\langle r_{cms}^{2}\rangle_{target}+\delta B \, .
\label{eq:11}
\endeq
The found small departure from the dipole size
dominated slope (\ref{eq:11}), $|\delta B|\lsim 1$\,GeV$^{-2}$,
is evidently related to the small
$R_{c}^{2} \approx 2\,{\rm GeV}^{-2} \ll {2\over 3}\langle
R_{ch}^{2} \rangle$. One can argue that
for the same reason the $r$ dependence of $B(\xi_{0},r)$
must be insensitive
to details of the infrared regularization.
The negligible $\delta B$ in
Eq.~(\ref{eq:11}) implies a simple boundary condition
$\eta(\xi_{0},r)\approx B(\xi_{0},0)\sigma(\xi_{0},r)$ and
an energy independent contribution $\approx B(\xi_{0},0)$
to the diffraction slope $B(\xi,r)$. Consequently,
the energy dependence of $B(\xi,r)$ and
$\alpha'_{eff}(\xi,r)$ are governed by the inhomogeneous term
$\beta(\xi,r)$ of Eq.~(\ref{eq:8}) and
only depend on the input dipole cross section
(\ref{eq:10}), which already has succesfully been
tested [15] against the DIS data.
The gluon-probed radius of the proton $R_{1}$ is a phenomenological
parameter to be determined from the experiment.
For the evaluation purposes, we approximate
$R_{1}$ by the proton charge radius $R_{ch}$:
\beq
B(\xi_{0},r \ll R_{c})\approx  {1\over 3}\langle R_{ch}^{2}
\rangle_{p}\approx 5.8{\, \rm GeV}^{-2}\, .
\label{eq:12}
\endeq

In Fig.~3 we present
the effective Regge slope $\alpha'_{eff}(\xi,r)$ which follows
from the solution of coupled Eqs.~(\ref{eq:3},\ref{eq:7},\ref{eq:8})
with the above described boundary conditions.
At $\xi \rightarrow \infty$,
$\alpha'_{eff}(\xi,r)$ tends to a $r$-independent
$\alpha_{\Pom}'=0.072$\,GeV$^{-2}$.
The very slow onset of the limiting value
$\alpha_{\Pom}'=0.072$\,GeV$^{-2}$
correlates nicely with the very slow onset
of the gBFKL asymptotics of $\sigma(\xi,r)$ and
of the proton structure function [3,8,15].
Slight oscillations in $\alpha_{eff}'(\xi,r)$
at small $\xi$ are a consequence of the oscillatory $r$-dependence
of large-$\nu$ eigenfunctions $E(\nu,r)$
(for the related discussion see [3]).

The above results for $\alpha'_{eff}(\xi,r)$ can be tested in
exclusive diffractive processes $\gamma^{*}p\rightarrow Vp$.
Ar fixed $\xi$, the dependence on the vector meson mass $m_{V}$
and the photon virtuality $Q^{2}$ is concentrated in
Eq.~(\ref{eq:2}) for the scanning radius $r_{S}$ and we predict
the same $B(\xi,r_{S})$ and $\alpha'_{eff}(\xi,r_{S})$ for all
reactions with the same $\xi$ and $r_{S}$ and/or $(m_{V}^{2}+
Q^{2})$, provided that
$r_{S}\lsim R_{V}$ [14,20].
For instance, $r_{S}(\Upsilon,Q^{2}=0)
\approx r_{S}(J/\Psi,Q^{2}\sim 120\,{\rm GeV}^{2}) \approx
r_{S}(\rho^{0}, Q^{2}\sim 200\,{\rm GeV}^{2}) \approx
0.13\,{\rm fm}$; $r_{S}(J/\Psi, Q^{2}\sim 30\,{\rm GeV}^{2})
\approx r_{S}(\rho^{0},Q^{2}\sim 60\,{\rm GeV}^{2})
\approx 0.2\,{\rm fm}$;  $r_{S}(J/\Psi,Q^{2}=0) \approx
r_{S}(\rho^{0},Q^{2}\sim 20\,{\rm GeV}^{2})\approx
0.4\,{\rm fm}$ and $r_{S}(\rho^{0}, Q^{2}\sim 3.5\,{\rm GeV}^{2})
\approx 0.76\,{\rm fm}$ (we refer to the dominant production
of the longitudinally polarized $\rho^{0}$ [14]).
The hatched areas in Fig.~3 indicate a variation of
$\alpha_{eff}'(\xi,r)$
over the HERA range of
{\sl c.m.s} energy $W=50-200$\,GeV (the higher values of
$\alpha_{eff}(\xi,r)$ correspond to a lower $W$). Remarkably,
at the HERA and lower energies, the subasymptotic
$\alpha_{eff}(\xi,r)\sim$(0.15-0.2)\,GeV$^{-2}$ is quite large,
close to $\alpha_{soft}'$ and by the factor 2-3 larger than
the asymptotic value $\alpha_{\Pom}'=0.072$\,GeV$^{-2}$.

The results for the diffraction slope in an approximation
(\ref{eq:12})
are presented in
Fig.~4. According to Eqs.~(\ref{eq:2},\ref{eq:11}), the diffraction
slope decreases with $Q^{2}$ and the vector meson mass.
The absolute value of $B(\xi,r)$ depends on the
assumed gluon probed radius of the proton $R_{1}$, whereas the rate of
the shrinkage $\alpha'_{eff}(\xi,r)$ does not.
The experimental determination
of $\langle R_{1}^{2}\rangle$ is of great
interest by its own.
For all the exclusive $\gamma^{*}p\rightarrow
Vp$ reactions we find a substantial rise of $B(\xi,r_{S})$, by
$\sim 1.5$\,GeV$^{-2}$, from the CERN/FNAL fixed target
range of $W\sim 10$\,GeV to the HERA collider energy
$W\sim 100$\,GeV. Here we have taken
$R_{c}=0.27$\,fm,
the experimental measurement of this
rise would be the best constraint on $R_{c}$.

The above analysis refers to the (infrared regularized)
perturbative gBFKL scattering
amplitude. For the description of semi-perturbative and soft
scattering phenomena, one needs to complement the gBFKL dipole
cross section by the soft nonperturbative, non-gBFKL, component
$\sigma^{(npt)}(\xi,r)$, which at $r\gsim R_{c}$ overwhelms
the above discussed perturbative gBFKL cross secrion
$\sigma^{(pt)}(\xi,r)$ [14,15].
This non-gBFKL soft cross section
has only a marginal effect on the proton
 structure function  at large
$Q^{2}$ [15], but contributes substantially to vector meson
production amplitudes unless $r_{S} \lsim R_{c}$ [14]. Of the above
$\gamma^{*}p\rightarrow Vp$ reactions,
only a real (and a virtual) photoproduction of the $\Upsilon$ and
a virtual photoproduction of the $J/\Psi$ at $Q^{2} \gsim 20
$\,GeV$^{2}$ and of the $\rho^{0}$ at $Q^{2}\gsim $40\,GeV$^{2}$ are
purely perturbative processes. For instance, in real
photoproduction of the $J/\Psi$, the
nonperturbative soft contribution makes $\sim 50\%$ of the
photoproduction amplitude.
Although the mechanism of this
nonperturbative soft interaction is not well undesrtood,
the geometrical
size dependence (\ref{eq:11}) could roughly be applicable
also to a soft production mechanism (for an example of the
nonperturbative model see [21]).
Furthermore, because
we find $\alpha'_{eff} \sim \alpha'_{soft}$, as a poor man's
approximation, we can use the above gBFKL evaluations of
the slope for the soft scattering too.
The estimates of
$B(\xi,r_{S})$ for a semi-perturbative real
photoproduction of the $J/\Psi$ and moderately virtual photoproduction
of the $\rho^{0}$, shown in Fig.~4, assume this poor man's
approximation.

The available data on the $J/\Psi$
photoproduction are not yet conclusive.
A direct selection of purely elastic events was performed only
in the FNAL E401 experiment [22] with the result
$B(J/\Psi;Q^{2}=0)=5.6 \pm
1.2$\,GeV$^{-2}$. The CERRN NMC result, $B(J/\Psi;Q^{2}=0)
=5.0 \pm 1.1$\,GeV$^{-2}$, comes
from a sample which contains background excitation of the
target proton [23]. These values of $B(J/\Psi;Q^{2}=0)$
for $W\sim 15$\,GeV
are consistent with our estimates shown in Fig.~4.
At the present stage of HERA experiments, a
direct rejection of proton excitation
is not yet possible [24,25]. The model-dependent
analysis of the first H1 data at
$\langle W\rangle =90$\,GeV gives estimates of
$B(J/\Psi;Q^{2}=0)$ ranging from 8.1 to 4.9 GeV$^{-2}$
[24], the first ZEUS data give
$B(J/\Psi;Q^{2}=0)=4.5\pm 1.4$\,GeV$^{-2}$ [25].
One needs the higher accuracy data from the both fixed target
FNAL/CERN and the collider HERA experiments.
Here one must bear in mind a well known rapid
rise of the diffraction slope towards
$t=0$ [26]. We calcuate
the diffraction slope at $t=0$, whereas the experimental data
on the vector meson production
correspond to a slope evaluated over quite a broad range
of $t$, typically up to $|t| \lsim 1$\,GeV$^{-2}$, which may
underestimate $B(V;Q^{2})$.
The ZEUS data on virtual photoproduction of the $\rho^{0}$
mesons give $B(\rho^{0};7<Q^{2}<25\,GeV^{2})=5.1+1.2-0.9(stat)
\pm 1.0(syst)$\,GeV$^{-2}$ [27], which
is close to the above cited $B(J/\Psi;Q^{2}=0)$,
 in agreement with our $(m_{V}^{2}+Q^{2})$
scaling. This result is also
consistent with
the predicted decrease of $B(\rho^{0};Q^{2})$
from real photoproduction in which
$B(\rho^{0};Q^{2}=0)= 10.6\pm0.4(stat)\pm 1.5(syst)$\,GeV$^{-2}$
[28].

To summarize, the purpose of the present paper has been an
exploration of the phenomenology of forward diffractive scattering
which emerges from the gBFKL dynamics. The intrusion of large
distance effects leads to a shrinkage of the diffraction cone
driven by the nonperturbative slope of the pomeron's Regge
trajectory $\alpha'_{\Pom}$. We presented the first evaluation
of the energy dependence of the gBFKL diffraction slope.
An
interesting finding is a large subasymptotic value of the
effective Regge slope $\alpha_{eff}'(\xi,r)$,
which is by the factor $\sim$(2-3) larger than
$\alpha_{\Pom}'$. For exclusive
production of all the vector mesons at HERA, we
find a substantial, by $\sim 1.5$\,GeV$^{-2}$, rise
 of the diffraction slope
$B(\gamma^{*}\rightarrow V)$ from
from the fixed target CERN/FNAL to the collider
HERA energy. From the scanning property in vector meson
production, we predict the scaling dependence
on the variable $(m_{V}^{2}+Q^{2})$,
which allows to relate diffracton slope for different
reactions.
Strong dependence of $\alpha_{\Pom}'$ on the gluon propagation
radius $R_{c}$ makes the rise of the diffraction slope
a sensitive probe of $R_{c}$.
\medskip\\
{\bf Acknowledgements:}
We are grateful to M.Arneodo, J.Nemchik and E.Predazzi for
helpful discussions.
B.G.Z. thanks S.Fantoni for hospitality extended
to him at SISSA, Trieste.
This work was partly supported by the INTAS grant 93-239
and by the grant NS9000 from the International Science Foundation.
\pagebreak\\

{\large \bf Figure captions:}\\

{\bf Fig.~1} - The Regge slope $\alpha_{\Pom}'$ of the pomeron
trajectory vs. the propagation radius of gluons $R_{c}$.
\\

{\bf Fig.~2} - The solid line shows the
diffraction slope $B(\xi_{0},r)$ for color
dipole-proton scattering in the two-gluon exchange Born approximation.
The dashed straight line shows the geometrical law
Eq.~(\ref{eq:11}) with.
The difference between the dashed line and
solid curve describes $\delta B$.\\

{\bf Fig.~3} - The energy ($\xi$) dependence of the effective Regge
slope $\alpha_{eff}'(\xi,r)$ vs. $\xi=\log(1/x_{eff})=
\log[W^{2}/(Q^{2}+m_{V}^{2})]$ changing by $\Delta \xi =2$
from $\xi=5.5$ (the curve $(a)$) to $\xi=13.5$ (the curve $(e)$).
The curve $(f)$ is for $\xi=17.5$. For reference, the curves
$(a),(b),(c),(d),(e)$ and $(f)$ correspond to $x_{eff}=W^{2}/(Q^{2}
+m_{V}^{2})=5\cdot 10^{-3},\,
7\cdot 10^{-4}\,, 9\cdot 10^{-5}\,, 1.25\cdot 10^{-5}\,,
1.6\cdot 10^{-7}$ and
$3\cdot 10^{-8}$, respectively. The horizontal line $(g)$ shows
the asymptotic value $\alpha_{\Pom}'=0.072$\,GeV$^{-2}$. The
hatched areas $A-D$ show the range of variation
of $\alpha_{eff}'(\xi,r_{S})$ over the HERA energy range
$W=$(50-200)\,GeV at a scanning radius $r_{S}$ relevant to the
reaction $\gamma^{*}p\rightarrow Vp$ at the photon's virtuality
$Q^{2}$: (A) $r_{S}=
0.12$\,fm, $\Upsilon(Q^{2}\sim 0)$, (B) $r_{S}=0.21$\,fm, $J/\Psi
(Q^{2}\sim 30\,{\rm GeV}^{2})$, (C) $r_{S}=0.4$\,fm, $J/\Psi(Q^{2}
\sim 0)$ and $\rho^{0}(Q^{2}\sim 20\,{\rm GeV}^{2})$, (D) $r_{S}=
0.76$\,fm, $\rho^{0}(Q^{2}\sim 3.5\,{\rm GeV}^{2})$.\\

{\bf Fig.~4} - The c.m.s. energy $W$ dependence of the
diffraction slope $B(\xi,r_{S})$
at a dipole size (scanning radius) $r_{S}$ relevant to different
diffractive $\gamma^{*}p\rightarrow Vp$ processes: (a) $r_{S}=
0.12$\,fm, $\Upsilon(Q^{2}\sim 0)$, (b) $r_{S}=0.21$\,fm, $J/\Psi
(Q^{2}\sim 30\,{\rm GeV}^{2})$, (c) $r_{S}=0.4$\,fm, $J/\Psi(Q^{2}
\sim 0)$ and $\rho^{0}(Q^{2}\sim 20\,{\rm GeV}^{2})$, (d) $r_{S}=
0.76$\,fm, $\rho^{0}(Q^{2}\sim 3.5\,{\rm GeV}^{2})$.


\begin{thebibliography}{99}

\bibitem{1} %
E.A.Kuraev, L.N.Lipatov and V.S.Fadin, {\sl Sov.Phys. JETP}
{\bf 44} (1976) 443; {\bf 45} (1977) 199;
Ya.Ya.Balitskii and L.N.Lipatov, {\sl Sov. J. Nucl. Phys.}
{\bf 28} (1978) 822.
L.N.Lipatov, {\sl Sov. Phys. JETP} {\bf 63} (1986) 904;
L.N.Lipatov. Pomeron in Quantum Chromodynamics. In: {\sl Perturbative
Quantum Chromodynamics}, editor A.H.Mueller, World Scientific, 1989.

\bibitem{2} %
J.~Bartels, {\sl J. Phys.} {\bf G19} (1993) 1611.

\bibitem{3} %
N.N.~Nikolaev, B.G.~Zakharov and V.R.~Zoller,
{\sl JETP Lett.} {\bf 59} (1994) 6;
{\sl Phys. Lett.} {\bf B328} (1994) 486;
{\sl J. Exp. Theor. Phys.} {\bf 78} (1994) 806;

\bibitem{4} %
N.N.~Nikolaev, B.G.~Zakharov and V.R.~Zoller,
{\sl JETP Lett.} {\bf 60} (1994) 694.

\bibitem{5} %
N.N.~Nikolaev and B.G.~Zakharov,
{\sl J. Exp. Theor. Phys.} {\bf 78} (1994) 598;
{\sl Z. Phys. } {\bf C64} (1994) 631;

\bibitem{6} %
F.Abe et al., {\sl Phys. Rev.} {\bf D50} (1994) 5519.

\bibitem{7} %
P.E.Volkovitski, A.M.Lapidus, V.I.Lisin and K.A.Ter-Martirosyan,
{\sl Sov.J.Nucl.Phys.} {\bf 24} (1976) 648;
A.Donnachie and P.V.L.Landshoff, {\sl Phys.Lett.} {\bf B123}
(1983) 345;
A.Capella, J.Tranh Van and J.Kwiecinski, {\sl Phys. Rev. Lett.}
{\bf 58} (1987) 2015;
B.Z.Kopeliovich, N.N.Nikolaev and
I.K.Potashnikova, {\sl Phys. Rev} {\bf D39} (1989) 769.

\bibitem{8} %
N.N.~Nikolaev and B.G.~Zakharov,
{\sl Phys. Lett.}
{\bf B327} (1994) 149.

\bibitem{9} %
R.E.Hancock and D.A.Ross,
{\bf Nucl. Phys.} {\bf B383} (1992) 575.

\bibitem{10} %
E.Levin, preprint {\bf TAUP-2221-94}.

\bibitem{11} %
V.N.~Gribov and L.N.~Lipatov, {\it Sov. J. Nucl. Phys.} {\bf 15} (1972)
438; L.N.~Lipatov, {\it Sov. J. Nucl. Phys.} {\bf 20} (1974) 181;
Yu.L.~Dokshitser, {\it Sov. Phys. JETP} {\bf 46} (1977) 641;
G.~Altarelli and G.~Parisi, {\it Nucl. Phys.} {\bf B126} (1977) 298.

\bibitem{12} %
M.Braun, preprint {\bf US-FT-17-94}.

\bibitem{13} %
N.N.Nikolaev, {\sl Comments on Nucl. Part. Phys.} {\bf 21} (1992) 41;
B.Z.Kopeliovich, J.Nemchick, N.N.Nikolaev and B.G.Zakharov,
{\sl Phys. Lett.} {\bf B309} (1993) 179.

\bibitem{14} %
J.~Nemchik, N.N.~Nikolaev and B.G.~Zakharov,
{\sl Phys. Lett.} {\bf B341} (1994) 228.

\bibitem{15} %
N.N.~Nikolaev and B.G.~Zakharov,
{\sl Phys. Lett.} {\bf B327} (1994) 157.

\bibitem{16} %
A.~Mueller and B.~Patel, {\it Nucl. Phys.} {\bf B 425} (1994) 471.

\bibitem{17} %
N.N.~Nikolaev and B.G.~Zakharov, {\it Z. Phys.} {\bf C49} (1991) 607;
{\bf C53} (1992) 331.

\bibitem{18} %
F.Low, {\sl Phys. Rev.} {\bf D12} (1975) 163; J.F.Gunion and
D.E.Soper, {\sl Phys. Rev.} {\bf D15} (1977) 2617.

\bibitem{19} %
B.G.Zakharov, {\sl Sov. J. Nucl. Phys.} {\bf 49} (1989) 860.

\bibitem{20} %
J.~Nemchik, N.N.~Nikolaev, E.Predazzi and B.G.~Zakharov, paper in
preparation.

\bibitem{21} %
J.R.Cudell, {\sl Nucl. Phys.} {\bf B336} (1990) 1.

\bibitem{22} %
M.Binkley et al., {\sl Phys. Rev. Lett.} {\bf 48} (1982) 73.

\bibitem{23} %
M.Arneodo et al., {\sl Phys. Lett.} {\bf B332} (1994) 195.

\bibitem{24} %
H1 Collab., T.Ahmed et al., {\sl Phys. Lett.} {\bf B338} (1994) 507.

\bibitem{25} %
ZEUS Collab., {\sl Phys. Lett.} {\bf B350} (1995) 120.

\bibitem{26} %
J.P.Burq et al., {\sl Phys. Lett.} {\bf B109} (1982) 111;
A.Schiz et al., {\sl Phys. Rev} {\bf D24} (1981) 26.

\bibitem{27} %
A.Whitfield, DIS-94 conference, Paris, April 1995.

\bibitem{28} %
M.Costa, Proceedings of International Conference on Deep
Inelastic Scattering and Related Subjects, Eilat, Israel,
6-11 February 1994, editor A.Levy, World Scientific, p.379.
\pagebreak\\
\end{thebibliography}
\end{document}